\documentclass[12pt,a4paper]{article}
\usepackage{a4wide}
\usepackage{latexsym}
\usepackage{epsf}
\usepackage{amssymb}

\begin{document}

\begin{flushright}
\small
IFT-UAM/CSIC-04-45\\
{\bf hep-th/0410252}\\
October $26$th, $2004$
\normalsize
\end{flushright}

\begin{center}


\vspace{2cm}

{\Large {\bf A Note on Supersymmetric G\"odel Black Holes, Strings and
    Rings of Minimal $d=5$ Supergravity}}

\vspace{2cm}


{\large Tom\'as Ort\'{\i}n}

\vspace{1cm}

{\it Instituto de F\'{\i}sica Te\'orica UAM/CSIC\\
  Facultad de Ciencias C-XVI,
  C.U. Cantoblanco,  E-28049-Madrid, Spain}\\
E-mail: {\tt Tomas.Ortin@cern.ch}

\vspace{3cm}


{\bf Abstract}

\end{center}

\begin{quotation}

\small

We show how any asymptotically flat supersymmetric solution of minimal
$d=5$ supergravity with flat base space $ds^{2}(\mathbb{E}^{4})$ can
be deformed into another supersymmetric asymptotically-G\"odel
solution and apply this procedure to the recently found supersymmetric
black-ring and black-string solutions. 

\end{quotation}

\newpage

\pagestyle{plain}


\textbf{1.} Black rings \cite{Emparan:2001wn} (black holes with
regular event horizons of topology $S^{2}\times S^{1}$) are
fascinating objects that challenge (in fact, violate), in higher
dimensions, the black hole uniqueness theorems that hold in four
dimensions, with potentially important consequences for entropy
calculations. Several generalizations of the original purely
gravitational solution of Ref.~\cite{Emparan:2001wn} have been found
to include electric charge \cite{Elvang:2003yy} and to make them
supersymmetric in the context of minimal and non-minimal $d=5$
supergravities
\cite{Elvang:2004rt,Gauntlett:2004wh,Elvang:2004ds,Bena:2004de,Behrndt:2004pn,Gauntlett:2004qy}.

In this note we describe yet another generalization of the
supersymmetric black rings of minimal $d=5$ supergravity to include
G\"odel asymptotics and preserving supersymmetry. We will show that
any asymptotically flat supersymmetric solution of minimal $d=5$
supergravity with flat base space it is possible to construct in a
systematic way another supersymmetric solution with G\"odel
asymptotics, and we will apply this procedure to the supersymmetric
BMPV black hole solution and to the supersymmetric black string
solution of Ref.~\cite{Bena:2004wv}. In the three cases
(supersymmetric black hole, string and ring) this deformation
preserves the geometry of the event horizon, but introduces closed
timelike curves (CTCs) far from the horizon.

We start by describing minimal $d=5$ supergravity and the
supersymmetric black ring solution of Ref.~\cite{Elvang:2004rt}.

\vspace{.3cm}

\textbf{2.} The action of the bosonic fields (metric and 1-form $V$)
of minimal $d=5$ supergravity \cite{Cremmer:1980gs} is, in our
conventions\footnote{Our conventions are those of
  Refs.~\cite{Lozano-Tellechea:2002pn,Meessen:2004mh}. In particular,
  we use mostly minus signature and our 1-form is $-2$ times that of
  Ref.~\cite{Gauntlett:2002nw}.}

\begin{equation}
\label{eq:N1d5SUGRAaction}
S= {\textstyle\frac{1}{16\pi G_{N}^{(5)}}}
\int  d^{5}x\sqrt{|g|}\,\left[R
-{\textstyle\frac{1}{4}}G^{2}
+{\textstyle\frac{1}{12\sqrt{3}}}
{\textstyle\frac{\epsilon}{\sqrt{|g|}}}
GGV\right]\, , 
\end{equation}

\noindent
where $G=dV$.


All the supersymmetric solutions of this theory can be classified
according to the timelike or null nature of the Killing vector
$V^{\mu}$, which is constructed as the bilinear of Killing spinors
$V^{\mu}=\bar{\epsilon}\gamma^{\mu}\epsilon$ \cite{Gauntlett:2002nw}.
We are interested in the former, since, in principle, they are the
ones that may describe stationary black holes and black rings. The
metric and 2-form field strength can always be brought to the form

\begin{eqnarray}
\label{eq:gensusysolution1}
ds^{2} & = & f^{2}(dt+\omega)^{2}-f^{-1}ds^{2}_{(4)}\, ,\\
&  & \nonumber \\
\label{eq:gensusysolution2}
G & = & -\sqrt{3}d[f(dt+\omega)] 
+{\textstyle\frac{2}{\sqrt{3}}}\mathcal{G}^{+}\, ,
\end{eqnarray}

\noindent
where the function $f$ and 1-form $\omega$ are independent of the time
coordinate $t$, $ds^{2}_{(4)}$ is a 4-dimensional hyper-K\"ahler
metric and $\mathcal{G}^{\pm}$ are the self- and anti-selfdual parts
of the 2-form $\mathcal{G}=fd\omega$ with respect to $ds^{2}_{(4)}$.
The scalar function $f$ and 1-form $\omega$ satisfy

\begin{eqnarray}
\label{eq:omegaequation}
d\mathcal{G}^{+} & = & 0\, ,\\
& & \nonumber \\
\label{eq:fequation}
\nabla_{\!(4)}^{2}f^{-1} -{\textstyle\frac{2}{9}} 
\mathcal{G}^{+}{}_{mn} \mathcal{G}^{+\, mn} & = & 0\, . 
\end{eqnarray}

Now we make the following observation: given a supersymmetric solution
of minimal $d=5$ supergravity, characterized by $f$ and $\omega$
satisfying Eqs.~(\ref{eq:omegaequation}) and (\ref{eq:fequation}), we
can always add to the 1-form $\omega$ a piece $W$ such that
$(dW)^{+}=0$.  The deformed $\omega^{\prime}=\omega +W$ satisfies
clearly Eq.~(\ref{eq:omegaequation}), and Eq.~(\ref{eq:fequation}) is
satisfied for the same $f$ since only $(d\omega)^{+}$ enters it. This
freedom is implicit in the wide class of solutions that we are going
to describe next.

\vspace{.3cm}

\textbf{3.} A particularly interesting class of solutions, including
those of interest for us (G\"odel spacetime and supersymmetric black
holes, strings and rings) can be obtained by using as $ds^{2}_{(4)}$
the 4-dimensional Euclidean space metric $ds^{2}(\mathbb{E}^{4})$,
written in the form of a Gibbons-Hawking instanton metric. These have
the general form

\begin{equation}
\label{eq:gensol1}
  \begin{array}{rcl}
ds^{2} & = & H d\vec{x}_{3}^{\, 2} 
+H^{-1} (d\psi +\chi)^{2}\, ,  \\
& & \\
d\chi & = & \star_{3}dH\, ,\\
\end{array}
\end{equation}

\noindent
where $ \star_{3}$ is the Hodge star in Euclidean space
$\mathbb{E}^{3}$ and $H$ is a harmonic function in $\mathbb{E}^{3}$.
The metric of $\mathbb{E}^{4}$ is recovered with the choices\footnote{If
  we use 3-dimensional spherical coordinates $r,\theta,\varphi$ so
\begin{equation}
ds^{2}(\mathbb{E}^{3})=
d\vec{x}_{3}^{\, 2}= dr^{2} +r^{2}d\Omega_{(2)}^{2}\, ,
\hspace{1cm}
d\Omega_{(2)}^{2} =d\theta^{2} +\sin{\theta}d\varphi^{2}\, .  
\end{equation}
then, if $H=1/|\vec{x}_{3}|$
\begin{equation}
\chi =\cos{\theta}d\varphi\, .  
\end{equation}
The coordinates $r,\theta,\varphi,\psi$ are related to the standard
4-dimensional spherical coordinates of $\mathbb{E}^{4}$
$\rho,\theta,\varphi,\psi$ in which
\begin{equation}
ds^{2}(\mathbb{E}^{4})= d\rho^{2} +\rho^{2}d\Omega^{2}_{(3)}\, ,
\hspace{1cm}
d\Omega^{2}_{(3)} = 
{\textstyle\frac{1}{4}}[d\Omega^{2}_{(2)} +(d\psi +\chi)^{2}]\, ,
\end{equation}
by
\begin{equation}
r=\rho^{2}/4\, .
\end{equation}
%
} 
$H=1$ or $H=1/|\vec{x}_{3}|$. Further one requires that
$\partial_{\psi}$ is a Killing vector of the full solution. It was
shown in Ref.~\cite{Gauntlett:2002nw} that, in these conditions, the
most general solution is specified by giving three functions $K,L $
and $M$ harmonic in the 3-dimensional Euclidean metric, in terms of
which

\begin{eqnarray}
\label{eq:gensol2}
f^{-1} & = & L+H^{-1}K  \, ,\\
& & \nonumber \\
\label{eq:gensol3}
\omega & \equiv & \omega_{5} (d\psi +\chi) +\hat{\omega}\, ,\\
& & \nonumber \\
\label{eq:gensol4}
\omega_{5} & = & M+ {\textstyle\frac{3}{2}}H^{-1}KL +H^{-2}K^{3}\, ,\\
& & \nonumber \\
\label{eq:gensol5}
\star_{3}d\hat{\omega} & = & HdM-MdH +{\textstyle\frac{3}{2}} (KdL-LdK)\, .
\end{eqnarray}

The freedom of modifying any solution by adding a piece $W$ with
$(dW)^{+}$ to $\omega$ is already contained in this form of the
solution, and is associated to the harmonic function $M$: it is easy
to see that the $M$-dependent part of $\omega$ in that solution

\begin{equation}
  \begin{array}{rcl}
W_{M} & = & M(d\psi+\chi) +\hat{\omega}_{M}\, ,\\
& & \\
\star_{3}d\hat{\omega}_{M} & = & HdM-MdH\, ,\\
\end{array}
\end{equation}

\noindent
has $(dW_{M})^{+}=0$, and it is due to this fact that $f$ does not
depend on $M$ at all. Thus, we can use the linearity of the Laplace
equation and the equation for $\hat{\omega}_{M}$ to modify $M$,
changing only the piece $W_{M}$ of $\omega$.

Observe that, when $K\propto H$, then $L$ enters the equation of $\omega$
in the same form as $M$ and the piece of $\omega$ that depends on it

\begin{equation}
  \begin{array}{rcl}
W_{L} & = & {\textstyle\frac{3}{2}}H^{-1}KL(d\psi+\chi) 
+\hat{\omega}_{L}\, ,\\
& & \\
\star_{3}d\hat{\omega}_{L} & = & {\textstyle\frac{3}{2}} (KdL-LdK)\, ,\\
\end{array}
\end{equation}

\noindent
also has $(dW_{L})^{+}=0$. Although $L$ appears in $f^{-1}$, now
$f^{-1}$ does not really depend on it, because we can add to it any
other harmonic function. In fact, all the solutions with $K\propto H$
can be rewritten as solutions with $K=0$ and two independent harmonic
functions $L,M$.

\vspace{.3cm}

\textbf{4.} In Ref.~\cite{Gauntlett:2004wh} it was shown that the
supersymmetric black ring of Ref.~\cite{Elvang:2004rt} was the case
associated to the choice

\begin{equation}
\label{eq:brsolution1}
\begin{array}{l}
{\displaystyle 
H = -\frac{1}{|\vec{x}_{3}|}\, ,\,\,\,\,\,\,\,\,
K = -\frac{q}{2|\vec{x}_{3}-\vec{x}_{3\, 1}|}\, ,\,\,\,\,\,\,\,\,
L = 1+\frac{Q-q^{2}}{4|\vec{x}_{3}-\vec{x}_{3\, 1}|}\, ,\,\,\,\,\,\,\,\,
}
\\
\\
{\displaystyle
M = \frac{3q}{4}
-\frac{3q |\vec{x}_{3\, 1}|}{4|\vec{x}_{3}-\vec{x}_{3\, 1}|}\, ,
\,\,\,\,\,\,\,\,
\vec{x}_{3\, 1} = (0,0,-R^{2}/4)\, ,
}
\\
\end{array}
\end{equation}

\noindent
which determines $\omega$ to be given by

\begin{equation}
\label{eq:brsolution2}
  \begin{array}{rcl}
\omega_{\psi}  \!\! \!\! & = &   \!\! \!\!
{\displaystyle
\frac{3q}{4}
\left[
1 -\frac{|\vec{x}_{3}|}{|\vec{x}_{3}-\vec{x}_{3\, 1}|}
-\frac{|\vec{x}_{3\, 1}|}{|\vec{x}_{3}-\vec{x}_{3\, 1}|}
\right]
-\frac{q|\vec{x}_{3}|}{16|\vec{x}_{3}-\vec{x}_{3\, 1}|^{2}}
\left[
3(Q-q^{2}) +\frac{2q^{2}|\vec{x}_{3}|}{|\vec{x}_{3}-\vec{x}_{3\, 1}|}
\right]
}\, ,\\
& & \\
\omega_{\varphi} \!\! \!\! & = &  \!\! \!\! \!\!
{\displaystyle
-\frac{3q}{4}
\left[
1 -\frac{|\vec{x}_{3}|}{|\vec{x}_{3}-\vec{x}_{3\, 1}|}
-\frac{|\vec{x}_{3\, 1}|}{|\vec{x}_{3}-\vec{x}_{3\, 1}|}
\right]
-\frac{q|\vec{x}_{3}|}{16|\vec{x}_{3}-\vec{x}_{3\, 1}|^{2}}
\cos{\theta} 
\left[
3(Q-q^{2}) +\frac{2q^{2}|\vec{x}_{3}|}{|\vec{x}_{3}-\vec{x}_{3\, 1}|}
\right]\, .
}
  \end{array}
\end{equation}

This asymptotically flat solution is characterized by four physical
parameters: mass $M$, angular momenta $J_{1},J_{2}$ and electric
charge $\mathcal{Q}$ given in terms of the three independent
parameters $Q,q,R$, by

\begin{equation}
M=\frac{3\pi}{4G_{N}^{(5)}}Q\, ,\,\,\,\,
J_{1} = \frac{\pi}{8G_{N}^{(5)}} q(3Q-q^{2})\, ,\,\,\,\,
J_{2} = \frac{\pi}{8G_{N}^{(5)}} q(6R^{2}+3Q-q^{2})\, ,\,\,\,\,
\mathcal{Q}=\frac{\sqrt{3}\pi}{G_{N}^{(5)}}Q\, .
\end{equation}

The BPS bound $M\geq \sqrt{3}|\mathcal{Q}|$ is saturated.  These
parameters are constrained by the condition $Q> q^{2} +2qR$ to avoid
CTCs. A 10-dimensional type~IIB supertube configuration that reduces
to the above solution of minimal $d=5$ SUGRA has been given in
Ref.~\cite{Elvang:2004ds} (see also Ref.~\cite{Bena:2004de}).  There
is an event horizon at $\vec{x}_{3}=\vec{x}_{3\, 1 }$ that has
topology and metric

\begin{equation}
ds^{2}_{(3)}= -\frac{q^{2}}{4}d\Omega_{(2)}^{2} 
-l^{2} d\psi^{2}\, . 
\end{equation}

\begin{equation}
l=\sqrt{3\left[\frac{(Q-q^{2})^{2}}{4q^{2}}-R^{2} \right]}\, .
\end{equation}

This solution has several different limits:

\textbf{(i)} $q=R=0$ corresponds to the extreme 5-dimensional
Reissner-Nordstr\"om black hole
$M-\sqrt{3}|\mathcal{Q}|=J_{1}=J_{2}=0$, which also belongs to the
class of solutions given in Eqs.~(\ref{eq:gensol1})-(\ref{eq:gensol5})
with $H=1/|\vec{x}_{3}|$ and is given by a single non-vanishing
harmonic function ($K=M=\omega=0$)

\begin{equation}
f^{-1}= L= 1 +\frac{Q}{4|\vec{x}_{3}|}=1+\frac{Q}{\rho^{2}}\, .
\end{equation}

The event horizon is placed at $\rho=0$ in these isotropic
coordinates, and its constant time sections have induced metric

\begin{equation}
ds_{(3)}^{2}=-Qd\Omega_{(3)}^{2}\, ,  
\end{equation}

\noindent 
and so they are round 3-spheres of radius $Q^{1/2}$, with
$A=2\pi^{2}Q^{3/2}$.

\textbf{(ii)} $R=0$ corresponds to the embedding in minimal $d=5$
supergravity of the charged, rotating, supersymmetric BMPV black hole
\cite{Breckenridge:1996is,Gauntlett:1998fz}
($M=\sqrt{3}|\mathcal{Q}|=\frac{3\pi}{4G_{N}^{(5)}}Q$,
$J_{1}=J_{2}={\displaystyle\frac{\pi}{8G_{N}^{(5)}} q(3Q-q^{2})}$)
\cite{Breckenridge:1996is}. This solution also belongs to the same
class, with $H=1/|\vec{x}_{3}|$, $K\propto H$ and $\hat{\omega}=0$,
and thus it can be written as a solution with $K=0$ and

\begin{equation}
  \begin{array}{rcl}
f^{-1} & = & L= 
{\displaystyle
1 +\frac{Q}{4|\vec{x}_{3}|}
}
\, ,\\
& & \\
\omega & = & W_{M} =M (d\psi +\chi)={\displaystyle \frac{J}{|\vec{x}_{3}|}}
(d\psi +\chi)\, .\\ 
\end{array}  
\end{equation}

Then one can see the supersymmetric BMPV black hole as a
supersymmetric deformation of the extreme Reissner-Nordstr\"om black
hole with $W=W_{M}$ \cite{Herdeiro:2002ft}.

This solution has a regular horizon at $\rho=0$ whenever $J<
\frac{1}{2}Q^{3/2}$. Its constant time sections have induced metric

\begin{equation}
ds_{(3)}^{2}=-\frac{Q}{4}(d\theta^{2}+\sin^{2}{\theta}d\psi^{2}) 
-\left[\frac{Q}{4}-\frac{4J^{2}}{Q^{2}}\right]
(d\varphi +\cos{\theta}d\psi )^{2}\, ,  
\end{equation}

\noindent 
and so they are squashed 3-spheres with
$A=2\pi^{2}\sqrt{Q^{3}-4J^{2}}$ \cite{Gauntlett:1998fz}.

\textbf{(iii)} The $R\rightarrow \infty$ limit gives the black string
solution of Ref.~\cite{Bena:2004wv}, which belongs to the same class
with


\begin{equation}
H=1\, ,\,\,\,\,
K=-\frac{q}{2|\vec{x}_{3}|}\, ,\,\,\,\,
L=1+\frac{Q}{2|\vec{x}_{3}|}\, ,\,\,\,\,
M=-\frac{3q}{4|\vec{x}_{3}|}\, ,\,\,\,\,
\end{equation}

\noindent
and, thus, it is given by

\begin{equation}
\label{eq:blackstringsolution}
  \begin{array}{rcl}
f^{-1} &= & {\displaystyle 1+\frac{Q}{|\vec{x}_{3}|} 
+\frac{q^{2}}{4|\vec{x}_{3}|^{2}}}\, ,  \\
& & \\
\omega & = & 
-{\displaystyle\left(\frac{3q}{2|\vec{x}_{3}|} 
+\frac{3qQ}{4|\vec{x}_{3}|^{2}}
+\frac{q^{3}}{8|\vec{x}_{3}|^{3}}\right)}d\psi\, ,\\
\end{array}
\end{equation}

\noindent
Actually, this solution is not quite a string\footnote{The
  supersymmetric string solution of minimal $d=5$ supergravity does
  not belong to the class of solutions we are studying, for which
  $V^{\mu}$ is null \cite{Gauntlett:2002nw}.}: when $q=0$, $\omega$
vanishes and the solution is an electrically charged extreme
Reissner-Nordstr\"om black hole smeared in the direction $\psi$, in
which it is not asymptotically flat. There is a horizon at
$\vec{x}_{3}=0$ with constant time slices of topology $S^{2}\times
\mathbb{R}$ and metric 

\begin{equation}
ds^{2}_{(3)}= -\frac{q^{2}}{4}d\Omega_{(2)}^{2} 
-3\frac{Q^{2}+q^{2}}{q^{2}} d\psi^{2}\, . 
\end{equation}

The $q=0,R\neq 0$ limit gives a solution with naked singularities.

\vspace{.3cm}

\textbf{5.} Now we want to introduce, via $W$s with $(dW)^{+}=0$
further deformations of the above solutions that preserve the
regularity of the horizons and, if possible, do not introduce other
pathologies like Dirac strings and CTCs. There seems to be very few,
or none at all, deformations with all these properties, and we are
going to settle deformations that do introduce CTCs but no Dirac
strings or naked singularities.

In Ref.~\cite{Herdeiro:2002ft} two particularly interesting $W$s
were considered\footnote{Our orientation is
  $\epsilon^{\theta\varphi\psi r}=+1$, so as to coincide with those of
  Refs.~\cite{Elvang:2004rt,Gauntlett:2004wh}, and seems to coincide
  with of Ref.~\cite{Herdeiro:2002ft}, which is not explicitly
  stated.}

\begin{eqnarray}
W_{B} & \equiv & {\displaystyle\frac{j}{\rho^{2}}} 
(d\psi +\cos{\theta} d\varphi)
\equiv {\displaystyle\frac{j}{\rho^{2}}} \sigma^{3}_{R}\, ,\\
& & \nonumber \\
W_{G} & \equiv & \lambda \rho^{2}  (d\varphi +\cos{\theta} d\psi)
\equiv  \lambda \rho^{2} \sigma^{3}_{L}\, .
\end{eqnarray}

$W_{B}$ is precisely the deformation that introduces rotation in the
extreme Reissner-Nordstr\"om black hole and is associated to a
harmonic function $M\sim 1/|\vec{x}_{3}|$. $W_{G}$ changes the
asymptotic behavior of the Reissner-Nordstr\"om black hole, which
still has a regular horizon basically because $W_{G}\sim \rho^{2}
\rightarrow 0$ in the near-horizon limit, but the solution now
asymptotes to a space of the same general form:

\begin{equation}
  \begin{array}{rcl}
ds^{2} & = & (dt+W_{G})^{2} - d\rho^{2} -\rho^{2}d\Omega^{2}_{(3)}\, ,\\
& & \\
G & = & -\sqrt{3}dW_{G}\, ,\\
  \end{array}
\end{equation}

\noindent
which is the celebrated maximally supersymmetric 5-dimensional analog
of G\"odel's solution \cite{Gauntlett:2002nw}. Thus, the deformed
solution is an asymptotically-G\"odel extreme Reissner-Nordstr\"om
black hole with an event horizon again placed at $\rho=0$ and with the
same geometry and area as in the undeformed black hole.

It is clear that we can use both deformations simultaneously, taking
$f$ as above and $\omega=W_{B}+W_{G}$, and obtain an
asymptotically-G\"odel BMPV black hole\footnote{A similar solution of
  (non-minimal) 5-dimensional supergravity has been given in
  Refs.~\cite{Herdeiro:2003un,Brecher:2003wq}. The metric is almost
  identical, the only difference being that $\sigma_{R}^{3}$ in
  $W_{G}$ is replaced by $\sigma_{L}^{3}$.}. To see this, we have to
prove that the surface $\rho=0$ has finite area and is null, but these
results follow immediately from the fact that $W_{G}$ vanishes fast
enough near the surface $\rho=0$. The area of the horizon will be
exactly that of the BMPV black hole. On the other hand, $W_{G}$ will
dominate the asymptotic behavior and, for large enough values of
$\rho$ there will be CTCs even in the regions in parameter space in
which the asymptotically flat BMPV black hole is free of them.
Finally, the physical parameters have to be redefined in accordance
with the new G\"odel asymptotics, that have a smaller isometry group.
This redefinition probably will not affect the mass, but may affect
the angular momenta which should be associated to the $SU(2)\times
U(1)$ part of the G\"odel isometry group, that may accommodate the two
independent angular momenta. A detailed study of this issue falls out
of the scope of this paper.

At this point it should be clear that any asymptotically flat
supersymmetric solution of minimal $d=5$ supergravity with flat base
space $ds^{2}(\mathbb{E}^{4})$ can be deformed into another
supersymmetric asymptotically-G\"odel solution using $W_{G}$. In
principle, though, we will have to check case by case if the
deformation changes the properties of the original solution, in
particular the values of charges and the existence and area of the
event horizon, although, as we have seen, this is highly unlikely. We
expect the presence of CTCs in all the asymptotically-G\"odel
solutions.

It is natural to perform a similar deformation of the
supersymmetric black ring solution with $W_{G}$. Now we have to check
explicitly that the deformed solution is in fact a black ring with a
regular horizon. We can do this by writing the metric in
$x\phi_{1},y,\phi_{2}$ coordinates\footnote{These are related to the
  spherical coordinates $\theta,\varphi,\psi,\rho$ by
\begin{equation}
\rho\sin{\theta/2} = \frac{R\sqrt{y^{2}-1}}{x-y}\, ,\,\,\,\,
\rho\cos{\theta/2} = \frac{R\sqrt{1-x^{2}}}{x-y}\, ,\,\,\,\,
\varphi=\phi_{1}-\phi_{2}\, ,\,\,\,\, 
\psi=\phi_{1}+\phi_{2}\, . 
\end{equation}
and, in this coordinate system
\begin{equation}
ds^{2}(\mathbb{E}^{4})=
\frac{R^{2}}{(x-y)^{2}}
\left[\frac{dy^{2}}{(y^{2}-1)} +(y^{2}-1)d\phi_{2}^{2}
+\frac{dx^{2}}{(1-x^{2})} +(1-x^{2})d\phi_{1}^{2}
\right]\, .  
\end{equation}
}, in which 

\begin{equation}
W_{G}=\frac{2\lambda R^{2}}{(x-y)^{2}}
\left[ (1-x^{2}) d\phi_{1} -(y^{2}-1)d\phi_{2}
\right]\, ,
\end{equation}

\noindent
redefining $y=-R/\bar{r}$ and taking the $\bar{r}\rightarrow 0$ limit
as in Ref.~\cite{Elvang:2004rt}. As the explicit calculation shows, in
the asymptotically flat case for constant $t$ and $\bar{r}=0$ we
obtain a regular metric for the horizon and in this calculation only
the terms singular in $\bar{r}$ contribute to this metric. Then, is
clear that $W_{G}$ will not modify the geometry and area of the
supersymmetric black ring event horizon. The regularity of the full
5-dimensional metric on the horizon is proven by analytical extension
of the solution, and this is not affected by the presence of $W_{G}$
either.  The same deformation can be used in the supersymmetric
concentric black ring solutions of Ref.~\cite{Gauntlett:2004wh}
without changing their near-horizon properties.

The $R=0$ limit of the new solution gives the G\"odel-BMPV black hole.
The $R\rightarrow \infty$ limit, though, gives again the same black
string solution as before. This is not surprising, since this limit is
also a sort of near-horizon limit and $W_{G}$ is irrelevant in that
limit. However, being supersymmetric, we are free to add directly to
the black string $W_{G}$ expressed in the new coordinates

\begin{equation}
W_{G}= 
\lambda(|\vec{x}_{3}|^{2}\sin{\theta}d\varphi -2|\vec{x}_{3}|d\psi )\, .  
\end{equation}

This term leads to G\"odel asymptotics in the directions orthogonal to
$\psi$ and does not change the geometry of the horizon since it
vanishes on it.

\vspace{.3cm}

\textbf{6.} If the very existence of the maximally superymmetric
G\"odel vacuum of minimal $d=5$ supergravity is already quite
remarkable, the possibility to ``excite'' this vacuum placing a large
variety of supersymmetric (and also non-supersymmetric
\cite{Gimon:2003ms}) objects such as black holes
\cite{Herdeiro:2002ft,Herdeiro:2003un,Brecher:2003wq} and black
strings \cite{Gimon:2003xk} in it is also very surprising. All these
solutions can be uplifted to solutions of 10-dimensional superstring
effective theories in which supersymmetric black rings will appear as
black superstubes in G\"odel-type or (T-dualizing) in $pp$-wave vacua.
Non-supersymmetric extensions of most of these solutions should also
exist. Work in this direction is in progress.

\vspace{1cm}

I would like to thank the High Energy Physics group of the University
of Barcelona for their hospitality during the last stages of this work
and in particular J.~Gomis and R.~Emparan for very helpful
conversations.  I am also indebted to M.M.~Fern\'andez for her
support. This work has been supported in part by the Spanish grant
BFM2003-01090.


\end{document}